\documentclass[figures,twocolumn]{revtex4}
\usepackage{graphicx}
\usepackage{bm,amsmath,amssymb}

\newcommand{\ud}{\mathrm{d}}
\newcommand{\e}{\varepsilon}
\newcommand{\eff}{\text{eff}}
\begin{document} 
\title{Granular Brownian Motor.} 
\author{B. Cleuren}
\affiliation{Hasselt University - B-3590 Diepenbeek, Belgium}
\author{C. \surname{Van den Broeck}}
\affiliation{Hasselt University - B-3590 Diepenbeek, Belgium}
\begin{abstract}
An asymmetric object, undergoing dissipative collisions with surrounding particles,  acquires 
a nonzero average velocity. The latter is calculated analytically by an expansion of the Boltzmann equation  and the result is compared with Monte Carlo  simulations.
\end{abstract}
\maketitle
\section{Introduction}
Brownian motors are spatially asymmetric constructions that, operating under nonequilibrium conditions, can rectify  thermal fluctuations. They have been the object of intense study over the past 15 years \cite{reimann2002}.  In this letter, we make the connection with another active field of research, namely  granular matter \cite{granular}. In such systems,  particles undergo dissipative colllisions and are, by construction, in nonequilibrium.  We therefore generically expect that a Brownian motor will arise if we 
break spatial symmetry in granular matter. 
To investigate this question in more detail, we propose here a minimal model, which has the advantage that the resulting systematic speed can be calculated exactly. We bypass the difficulties associated to the granular gas itself by assuming that the collisions between the gasparticles are elastic and that the gas is extremely diluted (ideal gas limit). Both spatial asymmetry and dissipation are introduced by considering an asymmetric object  which undergoes dissipative collisions with the surrounding particles. The outcome of our calculations is that the predicted speeds are quite large, namely comparable to the thermal speed, and should therefore be easily observable in experiment.

\section{The model}
\begin{figure}
\includegraphics[scale=0.7]{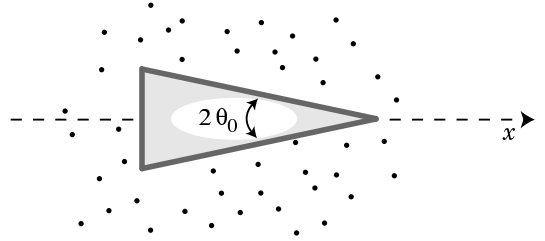}
\caption{Sketch of the system under 
consideration: an asymmetric object moving freely 
along the horizontal axis as a result of the 
inelastic collisions with the surrounding gas 
particles.}
\label{fig1}
\end{figure}

\begin{figure}
\includegraphics[scale=0.7]{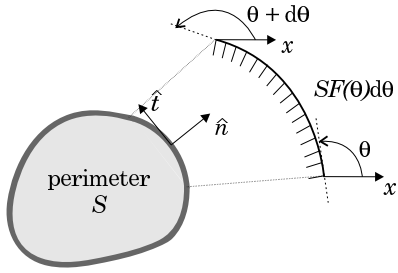}
\caption{Schematic representation of a convex object. Its shape is 
characterized by the shape probability function $F(\theta)$.}
\label{fig2}
\end{figure}

An asymmetric object, for example a triangle,  is free to move, however  without rotation, along a horizontal axis, see Fig.~\ref{fig1}. Its motion, with speed denoted by $V$, is induced by dissipative collisions with surrounding particles. For simplicity, we consider a two-dimensional system and assume that these particles are  an ideal gas initially at equilibrium at temperature $T$ in an infinitely large container. Hence the spatial distribution of the particles is uniform outside the object  and their velocity distribution is Maxwellian. Note that post-collisional particles are no longer at equilibrium. However, in the limit of infinite dilution, recollisions with the (convex) object do not take place and the hypothesis of molecular chaos (no correlations between the speed of the object and that of the particles prior to collisions) is valid. The following Boltzmann-Master equation is therefore a microscopically  exact starting point, describing the time evolution of the probabilitiy density $P(V,t)$:
\begin{multline}
\partial_t P(V,t)=\int_{-\infty}^{+\infty}\ud u 
\big[W(V-u;u)P(V-u,t)\\-W(V;u)P(V,t)\big].
\end{multline}
$W(V;u)$ is the transition probability per unit time that the object (mass $M$) changes speed 
from $V$ to $V+u$, due to collisions with the surrounding particles (mass $m$). Its explicit 
expression has to be derived from the laws of dissipative collisions, taking into account the 
geometrical configuration of the object. To remain consistent with the molecular chaos 
assumption, we restrict ourselves to convex objects. The latter can be fully characterized by 
the shape probability function $F(\theta)$, defined so that $F(\theta)\ud \theta$ is the fraction of the 
outer surface with polar angle between $\theta$ and $\theta +\ud \theta$ (see 
Fig.~\ref{fig2}). The total circumference will be denoted by $S$.
The object is constrained to move along a specific direction, say the $x$-axis, and its velocity is then $\vec{V}=V\vec{e}_{x}$. The velocity components of the particle are denoted by $v_x$ and $v_y$, and primes will be used to denote post-collisional speeds. Since there are no external forces in the $x$-direction, the corresponding component of the total momentum 
is conserved under collision:
\begin{equation}
mv_{x}'+MV'=mv_{x}+MV
\end{equation}
Furthermore, the interaction force is assumed to be orthogonal to the surface at the point of impact, implying that the tangential speed component of the impinging particle is also conserved:
\begin{equation}
\vec{v}'\cdot \hat{t}=\vec{v}\cdot \hat{t}
\end{equation}
Finally, the dissipative collision reduces the 
relative velocity orthogonal to the surface by a 
factor  $r$ ($0 \leq r \leq 1$), the so-called 
normal restitution coefficient:
\begin{equation}
(\vec{V}'-\vec{v}')\cdot \hat{n}=-r(\vec{V}-\vec{v})\cdot \hat{n}.
\end{equation}
The tangent and orthogonal unit vectors at the 
surface with polar angle $\theta$ are 
$\hat{t}=(\cos\theta, \sin\theta)$ and 
$\hat{n}=(\sin\theta,-\cos\theta)$ respectively.
The above equations determine uniquely the post-collisional speeds in terms of the pre-collisional ones. In particular, the change of speed $u$ of the object reads:
\begin{equation}
u=(1+r)\frac{m\sin^{2}\theta}{m\sin^{2}\theta+M}(v_{x}-V-v_{y}\cot 
\theta) \label{postV}
\end{equation}
The transition rate $W(V;u)$, which is the probability per unit time for such a change of speed, is thus given by:
\begin{widetext}
\begin{multline}
W(V;u)=\int_{0}^{2\pi}\ud \theta SF(\theta) 
\int_{-\infty}^{+\infty}\ud 
v_{x}\int_{-\infty}^{+\infty}\ud v_{y} 
\Theta\left[(\vec{V}-\vec{v})\cdot 
\hat{n}\right]\left\vert (\vec{V}-\vec{v})\cdot 
\hat{n}\right\vert  \\ \times \rho 
\phi(v_{x},v_{y})
\delta\left[u-(1+r)\frac{m\sin^{2}\theta}{m\sin^{2}\theta+M}(v_{x}-V-v_{y}\cot 
\theta)\right]
\end{multline}
where $\Theta\left[x\right]$ denotes the Heaviside function.  $\rho$ is the density of the gas and $\phi(v_{x},v_{y})$ the velocity distribution of the granular gas, taken to be a Maxwellian:
\begin{equation}
\phi(v_{x},v_{y})=\frac{m}{2\pi kT}\exp 
\left(-\frac{m(v_{x}^{2}+v_{y}^{2})}{2kT}\right)
\end{equation}
The integrals over the speed of the colliding gas particles can be performed explicitly, so that $W(V;u)$ reads:
\begin{multline}
W(V;u)=S\rho \sqrt{\frac{m}{2\pi 
kT}}\left(-u\Theta[-u]\int_{0}^{\pi}+u\Theta[u]\int_{\pi}^{2\pi}\right)\ud 
\theta F(\theta) \\ 
\times\frac{(m\sin^{2}\theta+M)^{2}}{(1+r)^{2}m^{2}\sin^{2}\theta}\exp 
\left[-\frac{m\sin^{2}\theta}{2kT}\left(V+\frac{u(m\sin^{2}\theta+M)}{(1+r)m\sin^{2}\theta}\right)^{2}\right]
\end{multline}
\end{widetext}
The remaining integral depends on the shape probability function of the object.
\section{Results}
The model introduced above can be considered as a granular variant of a thermal Brownian motor  \cite{vandenbroeck2004}. For the calculation of the resulting average drift velocity, we can proceed along lines 
similar to those described in \cite{meursPRE2004}, yielding an expansion for the stationary average drift velocity  in terms of the  small ratio $\e=\sqrt{m/M}$. The main steps of the calculation are  presented in the appendix. The resulting expressions for $\langle V \rangle$ and $\langle V^{2} \rangle$ read: 
\begin{widetext}
\begin{multline} \label{av}
\langle V \rangle = 
\sqrt{\frac{\pi}{2}}\sqrt{\frac{m}{M}}\sqrt{\frac{kT}{M}}\frac{1-r}{4}\frac{\langle 
\sin^{3}\theta \rangle}{\langle \sin^{2}\theta 
\rangle} 
-\left(\frac{m}{M}\right)^{3/2}\sqrt{\frac{kT}{M}\frac{\pi}{2}}\frac{1-r}{32}\left[\frac{\pi 
\langle \sin^{3}\theta \rangle^{3}}{\langle 
\sin^{2}\theta \rangle^{3}} \right.\\ \left. -2(5+r)\frac{\langle 
\sin^{3}\theta \rangle \langle \sin^{4}\theta 
\rangle}{\langle \sin^{2}\theta 
\rangle^{2}}+8\frac{ \langle \sin^{5}\theta 
\rangle}{\langle \sin^{2}\theta \rangle}\right] 
+\ldots
\end{multline}
and
\begin{equation} \label{av2}
\langle V^{2} \rangle 
=\frac{1+r}{2}\frac{kT}{M}+\frac{1}{16}\frac{kT}{M}\frac{m}{M}(1-r)\left[\pi 
\frac{\langle \sin^{3}\theta \rangle^{2}}{\langle 
\sin^{2}\theta \rangle^{2}}-4(1+r)\frac{\langle 
\sin^{4}\theta \rangle}{\langle \sin^{2}\theta 
\rangle}\right]+\ldots.
\end{equation}
\end{widetext}
The brackets involving the polar angle $\theta$ are averages with respect to the shape probability function  $F(\theta)$. In case of  elastic collisions, $r=1$, the object reaches a state of thermal equilibrium with the surrounding gas, that is $\langle V \rangle=0$ and $\langle V^{2} \rangle =kT/M$. For dissipative collisions, the object obviously does not equilibrate. To lowest order in $\e$, one verifies that the velocity of the object is still Maxwellian, but  at an (lower) effective temperature $T_{\eff}=(1+r)T/2$. As a result, there is a continuous flow of energy (heat) from the gas into the object. This energy flow, in conjuction with the spatial asymmetry, is the driving mechanims for its systematic motion. The direction of the motion is determined by the geometry. 
%%%%%%%%%%% Triangle
Turning to a specific example and for later comparison with the simulations, we consider an isocleses triangle with apex angle $2\theta_{0}$. The  shape probability function is:
\begin{multline}
F(\theta)=\frac{1}{2(1+\sin\theta_{0})}\Big(2\delta\left[\theta-3\pi/2\right]\sin\theta_{0}+\delta\left[\theta-\theta_{0}\right]\\+\delta\left[\theta-(\pi-\theta_{0})\right] 
\Big)
\end{multline}
and:
\begin{equation}
\langle \sin^{n}\theta \rangle 
=\frac{(-1)^{n}\sin\theta_{0}+\sin^{n}\theta_{0}}{1+\sin\theta_{0}}.
\end{equation}
The expressions for $\langle V \rangle$ and $\langle V^{2} \rangle$ now become: 
\begin{widetext}
\begin{multline}
\langle V \rangle = 
\sqrt{\frac{m}{M}}\frac{1-r}{4}\sqrt{\frac{kT}{M}\frac{\pi}{2}}(\sin \theta_{0}-1) 
-\left(\frac{m}{M}\right)^{3/2}\sqrt{\frac{kT}{M}\frac{\pi}{2}}\frac{1-r}{32}\bigg[\pi(\sin \theta_{0}-1)^{3}\\ -2(5+r)(\sin \theta_{0}-1)(1-\sin \theta_{0}+\sin^{2} \theta_{0})+8\frac{\sin^{4} \theta_{0}-1}{\sin \theta_{0}+1}\bigg] 
+\ldots
\end{multline}
and
\begin{equation}
\langle V^{2} \rangle 
=\frac{1+r}{2}\frac{kT}{M}+\frac{1}{16}\frac{kT}{M}\frac{m}{M}(1-r)\left[\pi 
(\sin \theta_{0}-1)^{2}-4(1+r)(1-\sin \theta_{0}+\sin^{2} \theta_{0})\right]+\ldots.
\end{equation}
\end{widetext}
We conclude that the triangle has an average negative speed: it always moves opposite to the direction in which it points ($\theta_{0}\in [0,\pi/2]$)! The average speed goes to zero, consistent with the fact that the motion is produced by fluctuations, when the mass ratio $m/M$ goes to zero. It becomes a maximum in the limit of high dissipation $r \rightarrow 0$ and maximum asymmetry $|\langle \sin^3\theta\rangle/\langle \sin^2\theta\rangle|=1$. 

\begin{figure}
\includegraphics[scale=0.75]{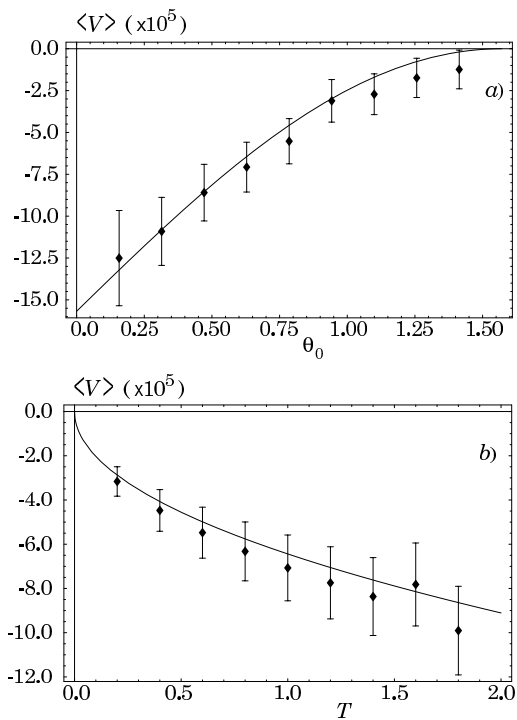}
\caption{Comparison between theory (full line) and simulation (dots) for the average velocity $\langle V \rangle$  of the triangle in a) as a function of the angle $\theta_{0}$ ($T=1.0$) and b) as a function of the temperature ($\theta_{0}=0.2\pi$). In both cases we set $m=1$, $M=100$, $\rho=0.00222$ and $r=0.95$.}
\label{fig3}
\end{figure}

For comparison with these theoretical results, we have performed stochastic simulations to generate trajectories of the Master equation. The highly efficient algorithm is based upon exact acceptance-rejection methods for
generating the Maxwellian inflow distribution \cite{garcia2006}; see the appendix C in \cite{meursPRE2004} for
a complete description of the algorithm. Averages were taken over 40 000 realizations, with each realization taking 100 000 time steps. The results are shown in Fig.~\ref{fig3}. Agreement between theory and simulations is excellent. Deviations are observed, as expected, when the mass ratio $m/M$ becomes smaller, but a surprisingly good agreement persists even for $M$ of the order of $m$, as shown in Fig.~\ref{fig4}.
\section{Discussion}
According to our theoretical result, the speed of the granular motor is of the order of the thermal speed $\sqrt{kT/M}$ and should therefore be easily observable. There are however a number of critical comments to be made concerning its experimental observation. 
First, one needs to address the fluctuations in velocity. Considering for simplicity the case of extreme dissipation and asymmetry, one finds $\langle V^2\rangle / \langle V\rangle^2= 32/\pi \times M/m + \ldots$.  We conclude that even for a mass ratio $m/M =.1$, r.m.s. speed  fluctuations are still 10 times larger than the average speed. Only for extremely light motors ($m/M=5$ according to the now less reliable prediction of the perturbative result) will fluctuations be of the same order as the average speed. This difficulty can of course be circumvented, either by taking enough sample trajectories, or by measuring the displacement for a time much longer than the correlation time of the velocity fluctuations. Second, the average speed is the result of a subtle unbalance between the collisions on different sides of the object. We therefore expect that the resulting systematic speed will depend sensitively on the velocity distribution that is used to describe the surrounding gas particles. For example, in the case of a driven granular gas, the velocity distribution can deviate strongly from the Maxwellian form and moreover depends significantly on the thermostat being used (see \cite{ernstJSP2006} for a recent review). Furthermore, a temperature anisotropy can appear in a granular gas because of the unidirectional energy input \cite{vandermeerEPL2006}. As a consequence, experimental observed velocities, using granular gases with a non-Maxwellian velocity distribution, are expected to yield significant deviations from the theoretical values predicted here. Deviations  from a Maxwellian distribtuion  can however be addressed by repeating the above calculations with the appropriate velocity distribution. Such calculations reveal that the average speed is typically of the same order of magnitude, although the detailed dependence on the shape of the object may be quite different. Finally, we note that the  discussion presented here is dealing with a two-dimensional system. The analysis can be reproduced with similar results for the case of three dimensions.

\acknowledgments
We thank R. Brito for communicating results on molecular dynamics simulations of the model presented here.

\appendix
\section*{Appendix}
Since we should recover equipartition in absence of dissipation, we expect that the kinetic energy of the motor is  of the order of $M\langle V^{2} \rangle \propto kT$. It is therefore convenient to switch to the dimensionless quantity $x$ of order $1$, defined as:
\begin{equation}
x=\sqrt{\frac{M}{kT}}V
\end{equation}
The Master equation is equivalent with the following set of coupled equations, describing the time evolution of the moments $\langle x^{n} 
\rangle=\int x^{n}P(x,t)\ud x$:
\begin{eqnarray}
\partial_t \langle x\rangle &=& \langle A_{1}(x)\rangle \label{onemoment}\\
\partial_t \langle x^{2} \rangle &=&2\langle
xA_{1}(x)\rangle +\langle A_{2}(x)\rangle \label{twomoment}\\
\partial_t \langle x^{3} \rangle &=& 3\langle x^{2}A_{1}(x)\rangle +3\langle xA_{2}(x)\rangle+\langle A_{3}(x) \rangle\\
\ldots && \nonumber
\end{eqnarray}
with $A_{n}(x)$ the so-called jump moment, defined as:
\begin{equation}
A_{n}(x)=\left(\sqrt{\frac{M}{kT}}\right)^{n}\int_{-\infty}^{+\infty}u^{n}W(\sqrt{kt/M}x;u)\ud u.
\end{equation}
\begin{figure}
\includegraphics[scale=0.75]{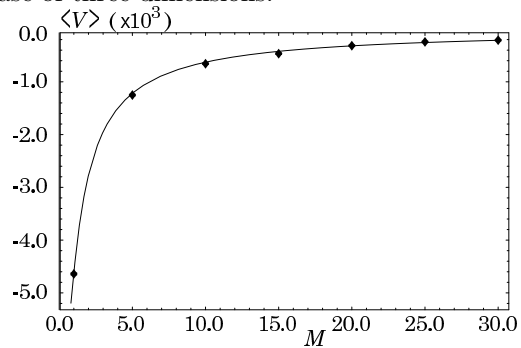}
\caption{Comparison between theory (full line) and simulation (dots) for the average velocity $\langle V \rangle$  of the triangle as a function of its mass $M$. Parameter values are set to $m=1$, $T=1.0$, $\theta_{0}=0.2\pi$, $\rho=0.00222$ and $r=0.95$.}
\label{fig4}
\end{figure}
Their explicit expression reads:
\begin{widetext}
\begin{multline}
A_{n}(x)=(-1)^{n}(1+r)^{n}2^{(n-1)/2}S\rho\sqrt{\frac{kT}{\pi 
m}}\e^{-n}\int_{0}^{2\pi}\ud \theta 
F(\theta)\left(\frac{\e^{2}\sin 
\theta}{1+\e^{2}\sin^{2}\theta}\right)^{n}e^{-\frac{\e^{2}x^{2}\sin^{2}\theta}{2}}\\ 
\times 
\left(\Gamma\left[1+\frac{n}{2}\right]\Phi\left[1+\frac{n}{2},\frac{1}{2},\frac{\e^{2}x^{2}\sin^{2}\theta}{2}\right]+\sqrt{2} 
\e x \sin\theta 
\Gamma\left[\frac{3+n}{2}\right]\Phi\left[\frac{3+n}{2},\frac{3}{2},\frac{\e^{2}x^{2}\sin^{2}\theta}{2}\right]\right).
\end{multline}
The function $\Phi\left[a,b,z\right]$ is the Kummer function. This new set of equations are fully coupled and equally difficult to solve as the original Master equation. However, the equations decouple in the limit $\e=\sqrt{m/M} \rightarrow 0$. Indeed, using the following series expansions:
\begin{eqnarray}
\Phi\left[1+\frac{n}{2},\frac{1}{2},z^{2}\right]&\approx& 
1+(2+n)z^{2}+\frac{1}{6}(8+6n+n^{2})z^{4}+... \\
\Phi\left[\frac{n+3}{2},\frac{3}{2},z^{2}\right]&\approx& 
1+\frac{1}{3}(3+n)z^{2}+\frac{1}{30}(15+8n+n^{2})z^{4}+...
\end{eqnarray}
one finds:
\begin{equation}
A_{1}(x)\approx(1+r)S\rho\sqrt{\frac{kT}{m}}\left\{-\sqrt{\frac{2}{\pi}}\langle 
\sin^{2}\theta \rangle x 
\e^{2}+\frac{1}{2}(1-x^{2})\langle \sin^{3}\theta 
\rangle \e^{3}+\frac{1}{3\sqrt{2\pi}}(6x-x^{3})\langle 
\sin^{4}\theta \rangle \e^{4}+...\right\}
\end{equation}
and:
\begin{equation}
A_{2}(x)\approx(1+r)^{2}S\rho\sqrt{\frac{kT}{m}}\left\{\sqrt{\frac{2}{\pi}}\langle 
\sin^{2}\theta \rangle \e^{2}+\frac{3}{2}x\langle 
\sin^{3}\theta \rangle \e^{3}+...\right\}
\end{equation}
\end{widetext}
The expressions  (\ref{av}) and (\ref{av2}) for $\langle V \rangle$ and $\langle V^{2} \rangle$ are now readily obtained, by solving the equations for the first two moments  $ \langle x\rangle$ and $ \langle x^2\rangle$ at the steady state up to order $\e ^{3}$.


\begin{thebibliography}{37}
\expandafter\ifx\csname natexlab\endcsname\relax\def\natexlab#1{#1}\fi
\expandafter\ifx\csname bibnamefont\endcsname\relax
  \def\bibnamefont#1{#1}\fi
\expandafter\ifx\csname bibfnamefont\endcsname\relax
  \def\bibfnamefont#1{#1}\fi
\expandafter\ifx\csname citenamefont\endcsname\relax
  \def\citenamefont#1{#1}\fi
\expandafter\ifx\csname url\endcsname\relax
  \def\url#1{\texttt{#1}}\fi
\expandafter\ifx\csname urlprefix\endcsname\relax\def\urlprefix{URL }\fi
\providecommand{\bibinfo}[2]{#2}
\providecommand{\eprint}[2][]{\url{#2}}

\bibitem[\citenamefont{Reimann}(2002)]{reimann2002}
\bibinfo{author}{\bibfnamefont{P.}~\bibnamefont{Reimann}}, \bibinfo{journal}{Phys. Rep.} \textbf{\bibinfo{volume}{361}},
  \bibinfo{pages}{57} (\bibinfo{year}{2002}).

\bibitem[{\citenamefont{Wolf}(1967)}]{granular}
\bibinfo{author}{\bibfnamefont{D.}~\bibnamefont{Wolf}},
\bibnamefont{and}
  \bibinfo{author}{\bibfnamefont{H.} \bibnamefont{Hinrichsen~(Editors)}},
  \emph{\bibinfo{title}{The physics of Granular Media}} (\bibinfo{publisher}{Wiley-VCH, Weinheim},
  \bibinfo{year}{2004}).

\bibitem[{\citenamefont{Ernst}(2006)}]{ernstJSP2006}
\bibinfo{author}{\bibfnamefont{M.~H.} \bibnamefont{Ernst}},
\bibinfo{author}{\bibfnamefont{E.} \bibnamefont{Trizac}}, \bibnamefont{and}
  \bibinfo{author}{\bibfnamefont{A.} \bibnamefont{Barrat}},
  \bibinfo{journal}{J. Stat. Phys.} \textbf{\bibinfo{volume}{124}},
  \bibinfo{pages}{549} (\bibinfo{year}{2006}).

\bibitem[{\citenamefont{Garcia}(2006)}]{vandermeerEPL2006}
\bibinfo{author}{\bibfnamefont{D.} \bibnamefont{van der Meer}}, \bibnamefont{and}
  \bibinfo{author}{\bibfnamefont{P.} \bibnamefont{Reimann}},
  \bibinfo{journal}{Eur. Phys. Lett.} \textbf{\bibinfo{volume}{74}},
  \bibinfo{pages}{384} (\bibinfo{year}{2006}).

\bibitem[{\citenamefont{VAn den Broeck}(2004)}]{vandenbroeck2004}
\bibinfo{author}{\bibfnamefont{C.}~\bibnamefont{Van den Broeck}},
\bibinfo{author}{\bibfnamefont{R.}~\bibnamefont{Kawai}},  \bibnamefont{and}
\bibinfo{author}{\bibfnamefont{P.}~\bibnamefont{Meurs}},
  \bibinfo{journal}{Phys. Rev. Lett.} \textbf{\bibinfo{volume}{93}},
  \bibinfo{pages}{090601} (\bibinfo{year}{2004}).

\bibitem[{\citenamefont{Meurs}(2004)}]{meursPRE2004}
\bibinfo{author}{\bibfnamefont{P.}~\bibnamefont{Meurs}},
\bibinfo{author}{\bibfnamefont{C.}~\bibnamefont{Van den Broeck}},
 \bibnamefont{and}
  \bibinfo{author}{\bibfnamefont{A.~L.} \bibnamefont{Garcia}},
  \bibinfo{journal}{Phys. Rev. E.} \textbf{\bibinfo{volume}{70}},
  \bibinfo{pages}{051109} (\bibinfo{year}{2004}).

\bibitem[{\citenamefont{Garcia}(2006)}]{garcia2006}
\bibinfo{author}{\bibfnamefont{A.~L.} \bibnamefont{Garcia}}, \bibnamefont{and}
  \bibinfo{author}{\bibfnamefont{W.} \bibnamefont{Wagner}},
  \bibinfo{journal}{J. Comp. Phys.} \textbf{\bibinfo{volume}{217}},
  \bibinfo{pages}{693} (\bibinfo{year}{2006}).

\end{thebibliography}
\end{document}